%% file: newfourd_paper.tex
\documentclass[11pt,a4paper,final]{IEEEtran}

\usepackage[utf8]{inputenc}				
\usepackage[T1]{fontenc} 				
\usepackage[german,british]{babel}		
\usepackage{textcomp}					
\usepackage{engord}					

\usepackage[protrusion,expansion=false,final]{microtype}		
\usepackage{ellipsis}										
\usepackage[style=numeric,sorting=none,backend=bibtex8]{biblatex}								
\usepackage[reqno,intlimits]{amsmath}		
\usepackage{amssymb}						
\usepackage{mathtools}					
\usepackage{siunitx}						
\usepackage{xfrac}							

\makeatletter
\adddialect\l@BRITISH\l@british
\makeatother

\usepackage{graphicx}
\graphicspath{{./g/}{./f/}{./m/report2/}{./bio/}}

\DeclareSIUnit{\dBm}{dBm}
\DeclareSIUnit{\dBc}{dBc}
\DeclareSIUnit{\Vpp}{Vpp}
\DeclareSIUnit{\bps}{bps}
\DeclareSIUnit{\baud}{Bd}
\DeclareSIUnit{\rad}{rad}

\providecommand{\shortexp}[1]{e^{#1}}			
\renewcommand{\iff}{\;\Longleftrightarrow\;}		

\providecommand{\Eptx}[2]{\mathrm{E}_{#1}\left[#2\right]}		

\providecommand{\vc}[1]{\mathbf{#1}}						
\providecommand{\vci}[2]{\mathbf{#1}_\mathrm{#2}}		
\providecommand{\sci}[2]{#1_\mathrm{#2}}				

\renewcommand{\j}{\num{j}}								

\providecommand{\hm}[1]{#1^\textnormal{H}}	

\begin{document}
\title{Four-dimensional signalling schemes -- Application to satellite communications}%
\author{Lionel Arend, Jens Krause, Michel Marso, Ray Sperber%
\thanks{L.~Arend was with SES S.A.~and the University of Luxembourg.}%
\thanks{J.~Krause and R.~Sperber are with SES S.A.}%
\thanks{M.~Marso is with the University of Luxembourg.}}%
\date{\today}
\maketitle
\begin{abstract}
In satellite communications both polarizations of an electromagnetic wave are used to transmit two separate signals. These two independent signals can be merged to form one dual-polarization, four-dimensional signal.

The present article pursues this idea and proposes different signal constellations to be used for four-dimensional signalling in satellite links. Analytical methods and simulations predict an increased power efficiency of these constellations with respect to currently used transmission methods. The cost of this advantage is evaluated considering the limited applicability in non-linear channels. 

Four-dimensional signalling also implies simultaneous reception on both polarizations. Such a combined reception allows the precision of timing and carrier recovery loops to be doubled. This claim is derived analytically and illustrated by simulating an example case.

An experimental transmitter/receiver pair was implemented and used to demonstrate a satellite transmission using  a four-dimensional, bi-orthogonal signal in the dual-polarization channel. The experimental verification confirms the presented simulation results.
\end{abstract}
%
\section{Introduction}
In a classical bandpass communication system, a single information carrying signal is modulated onto a carrier and fed to an antenna, cable, fibre or any other medium. Such a single carrier allows modulation in two dimensions: the in-phase and the quadrature amplitude. In four-dimensional modulation two bandpass channels are used, by exploiting for instance both polarizations that are available in free-space or in single-mode fibres. A first venture into four-dimensional modulation was proposed by Welti in \cite{Welti.1974}, showing different power-efficient constellations to be used in such a channel.


In 1991, a research group in Italy proposed to jointly modulate the orthogonal carriers in a fibre-link such as to more efficiently use all the resources of the channel \cite{Betti.1991}. They called the technique \mbox{\emph{4-Quadrature signalling}}. For quite some time, coherent signalling in the fibre channel remained a challenge, but with recent advances in high-speed digital circuits and fast analogue-to-digital converters, the optical world revived coherent, four-dimensional signalling \cite{Agrell.2009,ZhouXiang.2014}. Discussions on different four-dimensional symbol constellations and their expected power efficiencies have for instance been published in \cite{Karlsson.2012}. It seems that 4-D modulation has found its place in the optical community \cite{IvanB.Djordjevic.2014}. \cite{LotfollahBeygi.2014} even sums up proposals for coding schemes.

The main advantage of four-dimensional signalling is that symbol constellations with higher power efficiency may be used. In addition, \cite{IvanB.Djordjevic.2014} hints that non-linear signal degradations appearing on both polarizations could be efficiently equalized in a recursive fashion. Finally, 4-D signalling adds flexibility to bandwidth allocation, as wideband carriers might be split up between polarizations. In channels where crosspol (i.e.\ crosstalk between the polarizations) is an issue, demodulating both polarizations permits digital compensation measures \cite{Yamashita.2006}.

In many satellite applications, notably in broadcasting and broadband communications, the channel conditions are close to the ideal additive white Gaussian noise (AWGN) channel model. Fading or multipath propagation are more or less absent, crosspol is negligible for many practical applications. For the AWGN channel, the performance is well known and closely approached. As satellite transmission power is limited, a lot of effort is put into further optimizing  the power-efficiency by proposing different constellation layouts, respecting various constraints, for instance by considering the repeater non-linearity. For satellite broadcasting, a certain reference set of standard modulations is given by the DVB-S2 standard \cite{ETSI.200908}; examples for improvements include \cite{Candreva.2007}, \cite{Kayhan.2010} or \cite{ChunhuiQiu.2011}.

The satellite channel offers properties similar to a fibre. Many applications use highly directive, fixed antennas. These antennas have good polarization discrimination and are usually operated in a polarization-division multiplexing scheme. By that, the satellite channel offers as well the possibility for four-dimensional signalling using two orthogonal polarizations. A first effort to use four-dimensional signalling in satellite links was reported by Taricco \cite{Taricco.1993} in 1993. The approach is however insofar different from the one proposed in this article, as he used the quadrature-quadrature method from \cite{Saha.1989}, which increases the signal dimensionality by using four different basis functions on a single carrier signal.

The reception on two polarizations simultaneously can also offer advantages in receiver synchronization. In order to detect and de-map symbols from an incoming signal, a receiver first needs to synchronize to the signal. Carrier frequency and phase need to be estimated such as to provide an accurate local oscillator signal for downconversion. In a digital receiver, the symbol clock needs to be known, so that symbols can be extracted by downsampling at the right moment.

In four-dimensional modulation the carrier orthogonality is created through the state of polarization. This allows the transmission of two signals at \emph{exactly the same} carrier frequency, which is quite unique for free-space radio communications. Using a single local oscillator (LO) for frequency conversion, phase and frequency offset as well as phase noise would be equal on both carriers. This is used for instance by the \emph{Polarization Shift Keying} modulation \cite{Benedetto.1992} to enable direct detection. But also coherent modulation techniques should be able to exploit this property and enhance the carrier tracking fidelity. A similar consideration also holds for the symbol clock, which in 4-D modulation needs to be synchronous on both polarizations.

This article focuses on those two central aspects of 4-D signalling: Inspired by the findings in the optical community, it adapts the principles of four-dimensional modulation to satellite communications and discusses them as an alternative to PDM. A challenge is posed by the fact that orthogonal carriers are amplified separately by the satellite transponders. Results from coding theory are employed to construct power-efficient constellations that are subsequently compared by simulation with existing, classical satellite modulation methods. A demonstration transmitter and receiver pair is used to verify the simulations by experiment. It will be shown that satellite communications can indeed profit from the advantages of 4-D modulation, even though physical constraints limit its applicability in some cases. Possible solutions for circumventing the impact of high envelope variations are provided. Then, it is demonstrated how the tracking capabilities of a dual-polarization receiver can be improved. These findings are backed by accepted theoretical models as well as by the simulation of a specific example case.

The following section introduces a dual-polarization signal model and an accordingly modified definition of the signal-to-noise ratio (SNR). Section \ref{s_constellation} introduces the tools for constellation design, presents some examples and simulates their performance. Section \ref{s_synchro} expands existing theoretical gauges for tracking fidelity to 4-D modulation and illustrates them using a specific example. Finally Section \ref{s_concl} concludes with thoughts on the applicability of 4-D signalling and identifies some pointers for future work.
\section{Channel model}\label{s_chanmod}
In order to describe 4-D signalling, the common channel model for bandpass signalling \cite{Proakis.2008} is expanded to two conjointly modulated carrier signals. The signal consists of a series of symbols that are pulse-shaped and modulated into the target band. In classical coherent modulation, a particular symbol, numbered $i$, is defined by a complex scalar $\sci{c}{i}$ containing the carrier's amplitude and phase information.

As in 4-D signalling carriers on both polarizations are used, a pair of complex numbers $(\sci{c}{x,i},\sci{c}{y,i})$ is required to completely define a symbol. The projection of a symbol onto one of the polarizations X or Y, denoted $\sci{c}{x,i}$ and $\sci{c}{y,i}$, is from here on called symbol-element. The polarizations X or Y can be horizontal and vertical, right- or left-handed circular or any other pair of orthogonal states of polarization.

The modulated dual-polarization signal is represented in complex notation:
\begin{equation}
\begin{aligned}
\sci{s}{x}(t) & = \shortexp{\j\left[\left( \sci{\omega}{c} + \Delta\sci{\omega}{x\vphantom{y}}\right) t + \sci{\phi}{0,x}\right]}\sum_{i} \sci{c}{x,i} \cdot g\left(t-i\sci{T}{x}-\sci{\tau}{x\vphantom{y}}\right)\\
\sci{s}{y}(t) & = \shortexp{\j\left[\left( \sci{\omega}{c} + \Delta\sci{\omega}{y}\right) t + \sci{\phi}{0,y}\right]}\sum_{i} \sci{c}{y,i} \cdot g\left(t-i\sci{T}{y}-\sci{\tau}{y}\right)
\end{aligned}
\label{eq_sxsy_full}
\end{equation}
$g(t)$ is the baseband pulse-shape which, for convenience, represents the combined impulse response of transmission and reception filters. In radio-frequency signalling this is usually a raised-cosine pulse \cite{Proakis.2008}. The centre frequency is $\sci{\omega}{c}$. The physical signal is extracted by taking the real parts of $\sci{s}{x}(t)$ and $\sci{s}{y}(t)$.

The quantities $\Delta\omega$, $\sci{\phi}{0}$ and $\tau$ model carrier frequency offsets, initial phases and timing offsets \cite{Mengali.1997}. In general, they are different for both polarizations. Should some or all of these quantities be equal on both polarizations, a receiver can exploit this to improve its tracking capability; this is the subject of Section \ref{s_synchro} further below. For the first part, the analysis of 4-D signalling constellations, they do not need to be considered:
\begin{gather}
\Delta\sci{\omega}{x\vphantom{y}} = \Delta\sci{\omega}{y} = 0 \\
\sci{\phi}{0,x} = \sci{\phi}{0,y} = 0 \\
\sci{\tau}{x} = \sci{\tau}{y} = 0
\end{gather}
To have meaningful symbol de-mapping in four-dimensional signalling, it is important that the symbol clock frequency is equal on both polarizations. Some temporary differences might be compensated by buffers, such that the following equality must be fulfilled on average at least:
\begin{equation}
\sci{T}{x} = \sci{T}{y} = T
\end{equation}
The signal expression simplifies as follows and is comfortably expressed in vector form:
\begin{equation}
\begin{aligned}
\vc{s}(t) = \begin{pmatrix} \sci{s}{x}(t) \\ \sci{s}{y}(t) \end{pmatrix} & = \shortexp{\j\sci{\omega}{c} t} \sum_{i} \begin{pmatrix} \sci{c}{x,i} \\ \sci{c}{y,i} \end{pmatrix} \cdot g\left(t-iT\right)
\end{aligned}
\label{eq_sxsy_full}
\end{equation}
%
For the purpose of constellation analysis, it is sufficient to look at the symbol values alone. Assuming a properly downconverted and sampled signal, the symbols can be expressed using amplitude and phase or by their quadrature amplitudes (the index $i$ is dropped for brevity):
\begin{equation}
\vc{c} = \begin{pmatrix} \sci{c}{x} \\ \sci{c}{y} \end{pmatrix} = \begin{pmatrix} \sci{A}{x}\shortexp{\j\sci{\phi}{x}} \\ \sci{A}{y}\shortexp{\j\sci{\phi}{y}} \end{pmatrix} = \begin{pmatrix} \sci{x}{I}+\j\sci{x}{Q} \\ \sci{y}{I}+\j\sci{y}{Q} \end{pmatrix}\label{eq_cxcy}
\end{equation}
Because carriers on both polarizations are used, the signal occupies a bandwidth of twice the signalling rate, $B = \frac{2}{T}$. According to the \emph{Dimensionality Theorem}, the signalling scheme has thus $d = 2BT = 4$ dimensions \cite{Proakis.2008}. They correspond to the four quadrature amplitudes $\sci{x}{I}$, $\sci{x}{Q}$, $\sci{y}{I}$ and $\sci{y}{Q}$. For the case that X and Y are the linear horizontal and vertical polarizations, the symbol-vector $\vc{c}$ corresponds to the Jones-vector used in optics \cite{Hecht.1998}.
%
\subsection{Signal-to-noise ratio}
Choosing to express power using squared amplitudes, the average signal power $\sci{P}{s}$ and the corresponding average symbol energy $\sci{E}{s}$ are defined as follows:
\begin{equation}
\sci{P}{s} = \frac{1}{2\sci{T}{m}}\int_{\sci{t}{0}}^{\sci{t}{0}+\sci{T}{m}}\hm{\vc{s}(t)}\vc{s}(t)\,dt = \frac{\sci{E}{s}}{T}
\end{equation}
The superscript $\hm{\,}$ denotes a vector's hermitian, i.e.\ transposed and complex conjugate. The received signal $\vc{r}(t)$ is subject to thermal noise $\vc{n}(t)$, modelled as additive white Gaussian noise of double-sided power spectral density $\sci{N}{0}/2$:
\begin{equation}
\vc{r}(t) = \vc{s}(t) + \vc{n}(t)
\end{equation}
The noise power in the receiver is limited by the reception filters. Also noise is received on both polarizations; therefore twice the signalling rate has to be accounted for the noise bandwidth. Accordingly, the noise power $N$ will be:
\begin{equation}
N = B\sci{N}{0} = \frac{2}{T}\sci{N}{0}
\end{equation}
This leads to the following expression for the signal-to-noise ratio:
\begin{equation}
\text{SNR} = \frac{\sci{P}{s}}{N} = \frac{\sci{E}{s}}{T}\frac{T}{2\sci{N}{0}} = \frac{\sci{E}{s}}{2\sci{N}{0}}
\end{equation}
Interestingly, a factor \sfrac{1}{2} appears here, as opposed to the more common expression $\sci{E}{s}/\sci{N}{0}$. Although this might suggest that the SNR in a dual-polarization setup might be only half as large as in a classic channel, this coefficient is a consequence of the definition of signal energy as the sum energy on both polarizations.

Still, in order to achieve the same SNR as in a single-polarization setup, with the symbol rate constant, the dual-polarization signal has to be transmitted using twice the power inside twice the bandwidth. For this reason, also the bit-rates should be at least doubled with respect to a given classic modulation scheme.

In practice it is possible that the individual SNR values differ on each polarization. For the simulation and measurement results shown in this article, signal and noise power are always equally split between both polarizations, so that the SNR on each polarization equals the combined SNR.

For the results presented in the following sections to stay comparable to those in the literature, the expression for the signal-to-noise ratio is transformed to $\sci{E}{s}/\sci{N}{0}$ or $\sci{E}{b}/\sci{N}{0}$, according to the following formulas.
\begin{align}
\frac{\sci{E}{s}}{\sci{N}{0}} & = 2\cdot\text{SNR}\\
\frac{\sci{E}{b}}{\sci{N}{0}} & = \frac{2}{b}\cdot\text{SNR}
\end{align}
The variable $b$ stands for the number of bits transmitted per (dual-polarization) symbol.
\section{Four-dimensional signalling constellations}\label{s_constellation}
The increase from two to four dimensions re-opens the question of good symbol constellations for such a channel. Useful hints on how to approach constellation design come from the coding literature, where a high number of dimensions is not uncommon. The concepts of \emph{coding gain} and \emph{shaping gain} can orient the design and permit improvements with respect to classic I/Q modulations to be quantified. Good overviews are provided in \cite{Forney.1989} or \cite{Forney.1998} and, especially for the topic of shaping, \cite{Laroia.1994}. Constellation design in four dimensions seems to have been first described in \cite{Welti.1974}. The pertinent ideas can be summarized as follows:

An important figure for the noise sensitivity of a constellation is the minimum distance between signal points. It is given by the constellation packing and largely determines the symbol error probability when the signal-to-noise ratio is high. This minimum distance is given by the constellation packing. In a standard $2^{2n}$-point constellation for quadrature amplitude modulation ($2^{2n}$-QAM), the packing is a rectangular grid. In four dimensions, a more dense packing is possible, allowing more constellation points at equal mutual distance for the same average symbol energy. The densest packing in four dimensions is given by the chequerboard, or $\sci{D}{4}$, lattice \cite{Conway.1999}. The increase in power efficiency that can be achieved by choosing a more dense packing is called \emph{coding gain}.

Modulation schemes like $2^{2n}$-QAM have a rectangular shape, while the average symbol energy boundary, given by a radius $\sqrt{\sci{E}{s}}$ in the I/Q-plane, is \emph{circular}. For this reason cross-shaped constellations are proposed in \cite{Forney.1989} to fill the signalling space more efficiently. This principle can be extended to a higher number of dimensions, where it is then called \emph{constellation shaping}. In four dimensions the average symbol energy is bounded by a three-dimensional spherical surface, according to which the constellation space can be filled.

In short, the increase in dimensions allows constellation points to be packed more densely, which in practice means that more information can be transmitted using the same power at the same symbol error rate (SER). In addition, the signal space defined by a power constraint can be filled up more efficiently. 

These principles from coding theory can be used in physical modulation as well to increase the power efficiency in uncoded modulation. They are used in the following sections to design four-dimensional constellations. Different categories of constellations are proposed, starting with lattice-based constellations for the linear channel, which illustrate the gross signalling gain achievable from the dimensional increase. Then, the degrees of freedom are successively reduced to yield designs that are more appropriate for the non-linear channel available in many satellite communication links. Simulation results are presented for the different constellations in the AWGN channel, showing their symbol error rate as a function of the SNR. 

Considering the bandwidth and power use of dual-polarization modulation, the 4-D modulation schemes have to be compared with classical I/Q modulation techniques in dual operation (e.g.\ quaternary phase-shift-keying (QPSK) simultaneously on both polarizations). This ensures a fair comparison between single-polarization and dual-polarization techniques, because spectrum usage and transmission rates are equal. The SER of the classical modulation schemes is derived from known error probability formulas \cite{matlabsoft}. For a dually operated technique to lead to a correct decision, \emph{both} underlying symbols need to be detected correctly. In consequence, the error probability $\sci{P}{c}$ in dual operation has to be adjusted accordingly:
\begin{align}
\sci{P}{c,\text{dual}} & = \sci{P}{c,\text{single}}\cdot\sci{P}{c,\text{single}} \notag \\
\iff  1-\sci{P}{e,\text{dual}} & = \left(1-\sci{P}{e,\text{single}}\right)\cdot\left(1-\sci{P}{e,\text{single}}\right) \notag \\
\iff \sci{P}{e,\text{dual}} & =  2\sci{P}{e,\text{single}} - \sci{P}{e,\text{single}}^2
\end{align}
Probabilities subscripted with c are for correct decisions, while e stands for errors.

For the simulation results shown below, the error rates of various signalling constellations are compared by the SNR required to decrease the error rate below the threshold of \num{1e-4}. This is estimated to be a good compromise between the achievable asymptotic result at high SNR and the results for a practical operation scenario at lower SNR, where coding techniques are used for error correction.
\subsection{Lattice-based constellations}\label{ss_lam}
Targeting maximally power-efficient signalling, the first set of constellations is based on the $\sci{D}{4}$-lattice. To this end, a large sample of $\sci{D}{4}$-lattice is generated and positioned in space such as to free the origin from a node. The constellation is then carved out by delimiting the lattice by a spherical boundary centred around the origin \cite{Laroia.1994}. The radius of the sphere can be freely chosen such as to include the desired number of signal points. In this way, a maximally dense and shaped constellation is created with minimum average power. The lattice itself provides for the \emph{coding gain}, the spherical cut-out for the \emph{shaping gain}.

The so-created constellations are referred to as \emph{lattice amplitude modulation} (LAM) constellations in this article. LAM constellations can be viewed as four-dimensional correspondents to classic, rectangular QAM, or to the cross-shaped constellations from \cite{Forney.1989}. An 88-LAM and 256-LAM constellation have been generated. In spectral efficiency, the latter corresponds to a dual 16--APSK or 16-QAM constellation, the former approximately to dual 8-PSK or 8-QAM\footnote{The generation procedure does not necessarily create constellations with a number of points being a power of 2. It was not possible to elegantly arrive at 64-points, so the choice falls on the next larger constellation with 88 lattice nodes.}. Both generated LAM constellations have already been published in \cite{Welti.1974} and are thus known; they have not been considered as an alternative to the polarization division multiplex or been compared with other, classical techniques in a dual-polarization setup.

\begin{figure*}
\centering
\resizebox{0.75\textwidth}{!}{\includegraphics{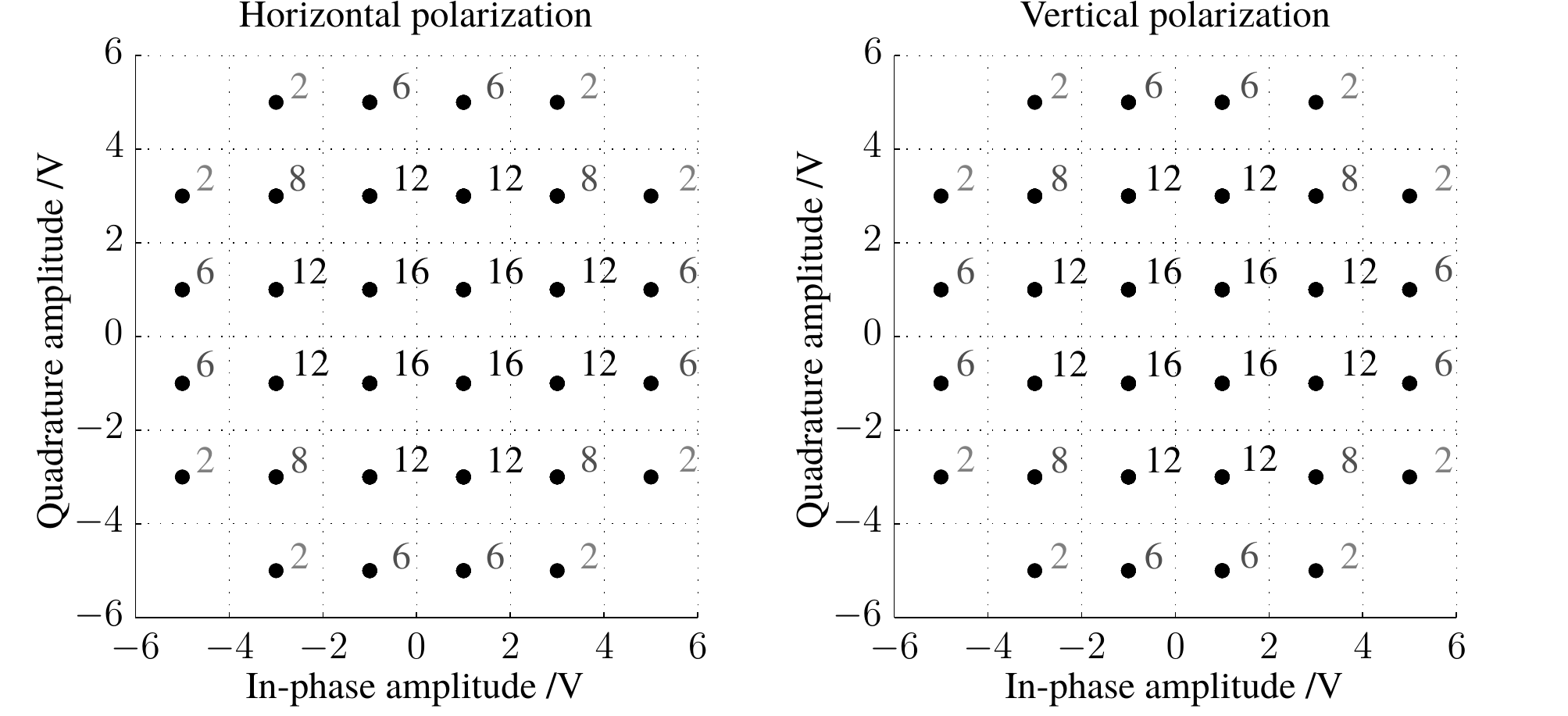}}
\caption{256-LAM constellation projected onto both I/Q planes. Numbers indicate the symbol-elements' frequency if every symbol in the constellation is selected once.}
\label{plot_256const}
\end{figure*}

Orthogonal projections\footnote{These are called \emph{constituent 2-D constellations} in the coding literature.} of the 256-LAM constellation onto each polarization are displayed in Figure \ref{plot_256const}. The numbers indicate how many symbols will resolve to a given symbol-element in either projection. They symbolize the probability that a certain symbol-element will be selected when all symbols are equiprobable. Note that different orthogonal projections are possible. The present one is chosen such that the number of symbol-elements is minimized.

Figure \ref{plot_4DmodsLAM} shows a symbol error rate plot for the LAM constellation compared with reference modulation schemes. Simulation results and calculated union bounds (UB) \cite{Conway.1999} are given. Both LAM constellations show considerable advantages over their classical correspondents. At the error rate of \num{1e-4}, the 88-point constellation gains \SI{2,2}{\deci\bel} with respect to dual 8-PSK. This comparison is difficult to justify however, as 8-PSK is a constant-power modulation. For this reason another comparison is proposed with a hexagonally-arranged 8-QAM constellation. The hexagonal 8-QAM constellation is more advantageous than a rectangular 8-QAM, by its energy efficiency and constellation peak-to-average power ratio. Its coordinates are shown in Appendix \ref{apx_hexqam}. Comparing it with the proposed 88-point LAM, the power advantage shrinks to \SI{0,7}{\deci\bel}, which can be considered the effective advantage achieved by the dimensional increase.
\begin{figure}
\centering
\resizebox{\linewidth}{!}{\includegraphics{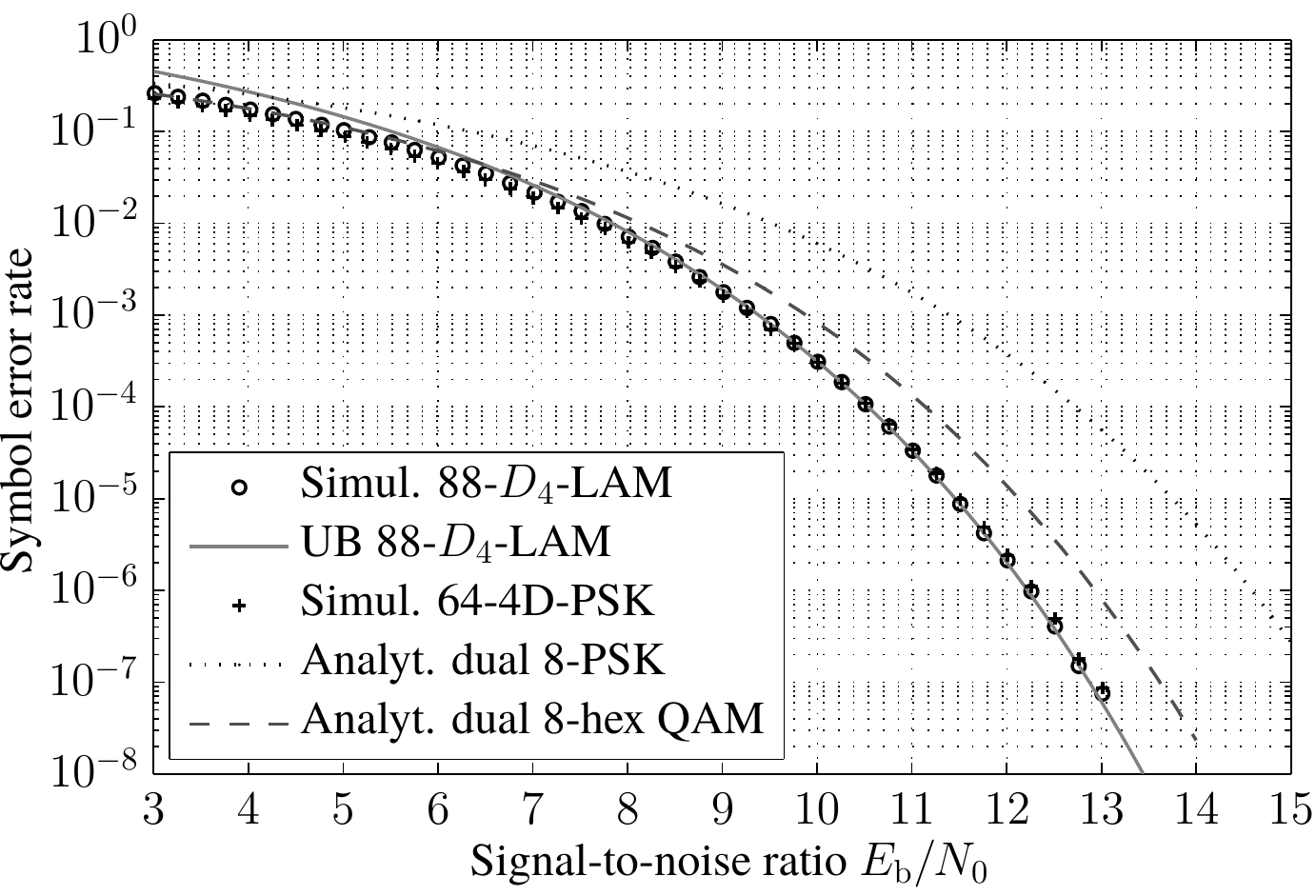}}
\caption{SER as a function of SNR for different 6 bit per symbol 4-D constellations in the AWGN channel}
\label{plot_4DmodsLAM}
\end{figure}
When comparing the techniques with a spectral efficiency of 8 bit per symbol, the 256-point LAM constellation outperforms 16-QAM by \SI{1,3}{\deci\bel} and 16-APSK\footnote{Relationship between inner and outer threshold: \num{2,5}. By extrapolating the proposed radii in the DVB-S2 standard, this can be estimated to be optimal relationship for uncoded 16-APSK.} by \SI{1,5}{\deci\bel}. The comparison with 16-QAM is considered more appropriate, because the reduced power efficiency for APSK is the result of a placement optimized for the non-linear satellite channel; the arrangement no longer follows the structure of a lattice. 
\begin{figure}
\centering
\resizebox{\linewidth}{!}{\includegraphics{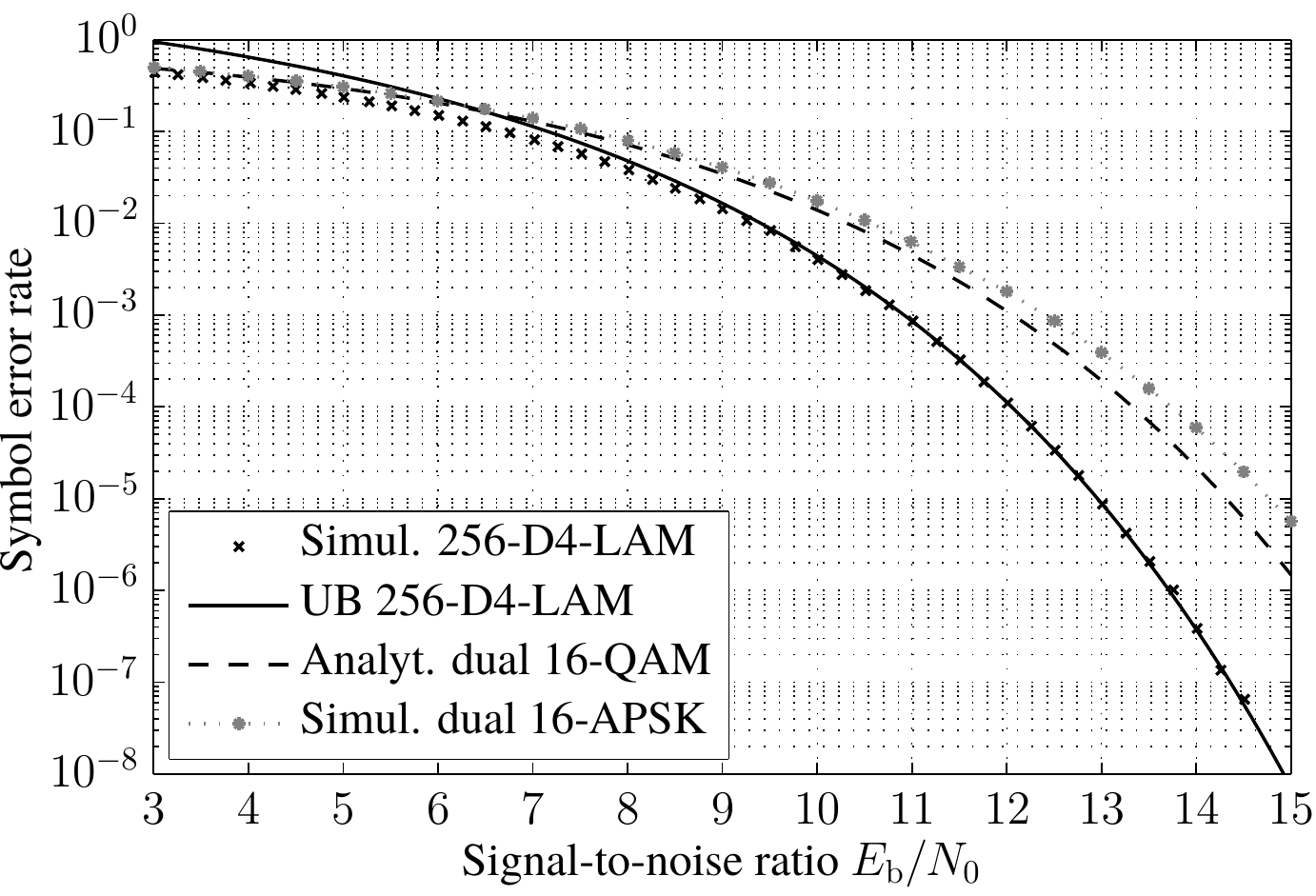}}
\caption{SER as a function of SNR for different 8 bit per symbol 4-D constellations in the AWGN channel}
\label{plot_4DmodsLAM}
\end{figure}
For both LAM constellations, a significant signalling gain is achieved, demonstrating the possibility of increasing the power efficiency by using both polarizations. The impact of using a four-dimensional signalling technique is more important at higher spectral efficiencies; more points can be placed and thus more degrees of freedom for an optimal shaping are available.
\subsection{Sphere-based constellations}\label{ss_sphereconst}
Phase shift keying (PSK) or amplitude and phase shift keying (APSK) are used for transmissions with amplifiers at or close to saturation. Constellations like QPSK and 8-PSK have all constellation points on an equal-power circle in the I/Q-plane. In this way, envelope variations are kept at a minimum and distortion small.

To construct similar constellations in four-dimensions, one can place a fixed number of points equidistantly onto the three-dimensional surface of a sphere in 4-D space; then all the symbols have equal energy. These are the constellations initially proposed in \cite{Betti.1991}.

A 64-point constellation is used, denoted 64-4D-PSK, to be compared with dual 8-PSK. The exact arrangement of points is taken from Sloane's website on spherical packings \cite{N.J.A.Sloane.}. The SER-to-SNR curve is shown in the same Figure \ref{plot_4DmodsLAM} as for the LAM constellations.

Interestingly, the simulated SER is very close to that of the 88-point LAM constellation and thus offers a similar advantage compared to classical modulation techniques.

As the constellation is the result of a numerical optimization, it has a quite chaotic structure and is therefore not shown. More \emph{orderly} constellations might be created using the algorithm described by Leopardi in \cite{PaulLeopardi.2007}, at the price of some power efficiency.

The constant-power limitation in 64-4D-PSK is effective for both polarizations together. This is only meaningful when both polarizations are processed and amplified conjointly as well. As contemporary satellite amplifiers have to amplify both polarizations separately, they will be subject to the envelope variations on each carrier. Thus, for the proposed 64-4D-PSK constellation, amplifiers would require some back-off.
%
\subsection{Cylinder-based constellations}
If signal envelope variations are to be completely avoided also on individual polarizations, the constraints on constellation design have to be further restricted. By fixing the power on both carriers, only two degrees of freedom remain: the phases. The same principle of dense lattices, used above in the \emph{quadrature amplitude space}, can be applied inside the two-dimensional \emph{phase space}. As in two dimensions the hexagonal grid is the most dense, a constellation can be generated by laying out points accordingly. The constellation layout can be imagined as intersecting cylinders in 4-D space, hence the classification as \emph{cylinder-based}.

\begin{figure}
\centering
\resizebox{0.71\linewidth}{!}{\includegraphics{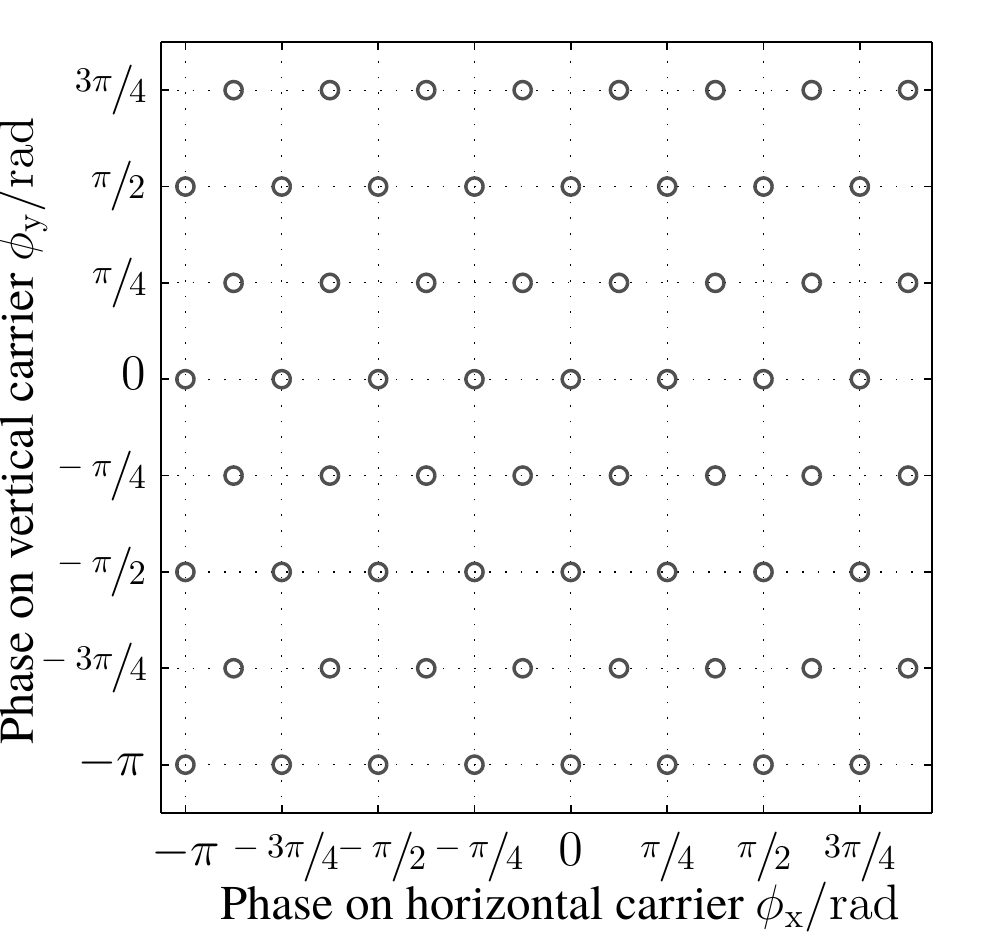}}
\caption{Phase diagram of 64-point hexagonally-arranged 4-D PSK}
\label{plot_cylphase}
\end{figure}

The phase space is of quadratic size, $[0;2\pi[\times[0;2\pi[$ and a closed surface (i.e.~The boundaries are \emph{glued} together); whereas the hexagonal grid has a non-rational side-ratio: A quadratic chunk can only be created approximately, for a very high number of points. If 64 points are used, the lattice needs to be slightly squeezed in order to \emph{fit}; this deformation costs power efficiency. A constellation plot in the two-dimensional \emph{phase space} is shown in Figure \ref{plot_cylphase}.

\begin{figure}
\centering
\resizebox{\linewidth}{!}{\includegraphics{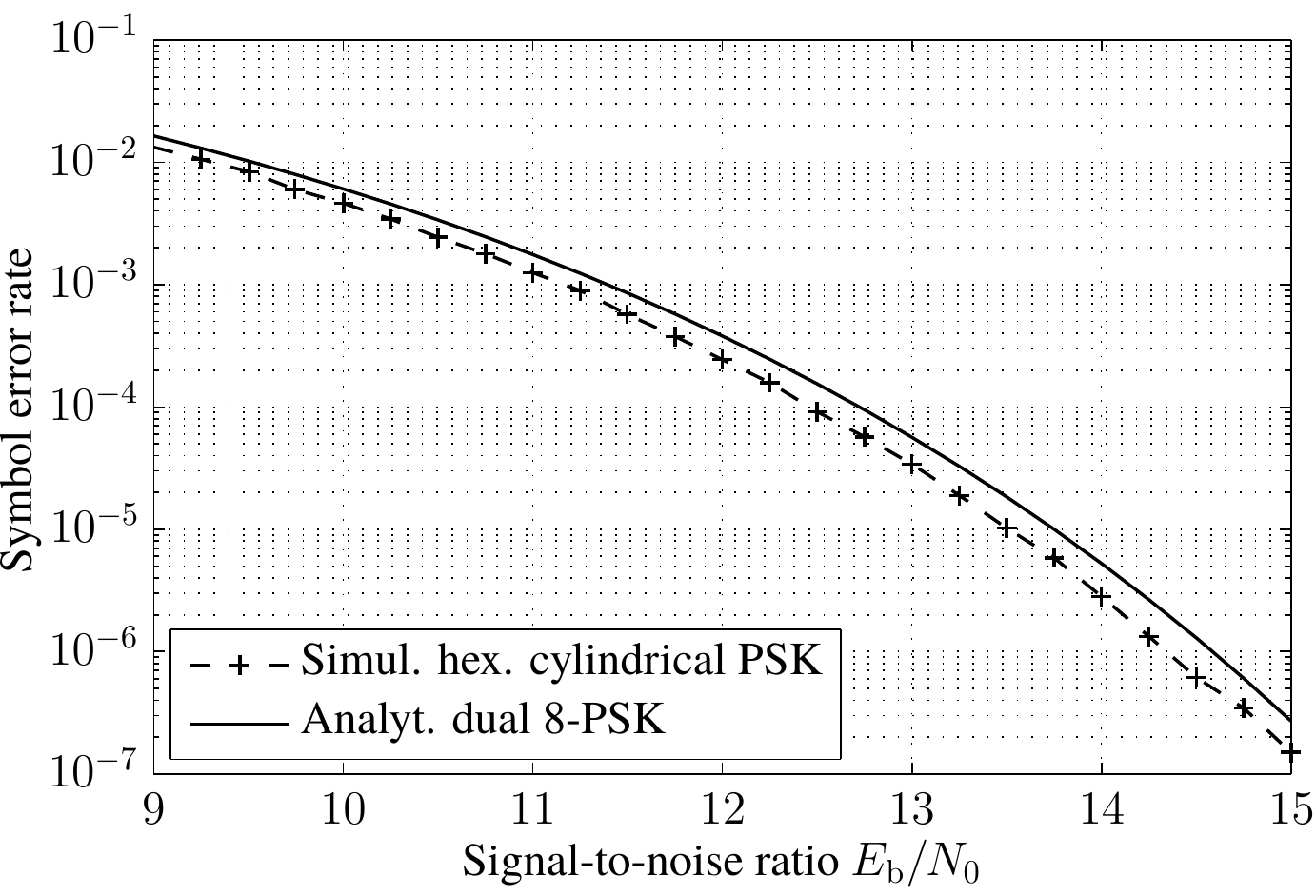}}
\caption{SER as a function of SNR comparing dual 8-PSK with 64-point 4-D PSK}
\label{plot_4DmodsPSK}
\end{figure}

This time, the comparison with 8-PSK is fully justified and shown in Figure \ref{plot_4DmodsPSK}. The gain in SNR is quite small with only \SI{0,2}{\deci\bel}. Still, it has to be considered that this gain is achieved solely by rearranging phase values in a constant-amplitude constellation. It remains open whether this gain is still valid when bit-error rates are compared. 

Taking APSK as a model, one can envisage similar constellations with multiple cylindrical shells. The construction of such constellations seems to be more involved and probably best undertaken with the help of numerical optimization procedures. An algorithm like that proposed in \cite{Candreva.2007} should be extendible to four-dimensional signals. An extensive search for APSK-like four-dimensional constellations was not considered in the scope of the present work.
\subsection{Bi-orthogonal constellation}\label{ss_biortho}
A recent publication in optical communications, comparing signal constellations in two, three and four dimensions, identifies four-dimensional, bi-orthogonal signalling as the most spectrum efficient of all, even beating binary PSK (BPSK) \cite{Karlsson.2009}. The bi-orthogonal modulation is a constant energy modulation, consisting of the symbol points $\{\num{-1};\num[retain-explicit-plus]{+1}\}$ on each of the four orthogonal axes.

\begin{figure}
\centering
\resizebox{\linewidth}{!}{\includegraphics{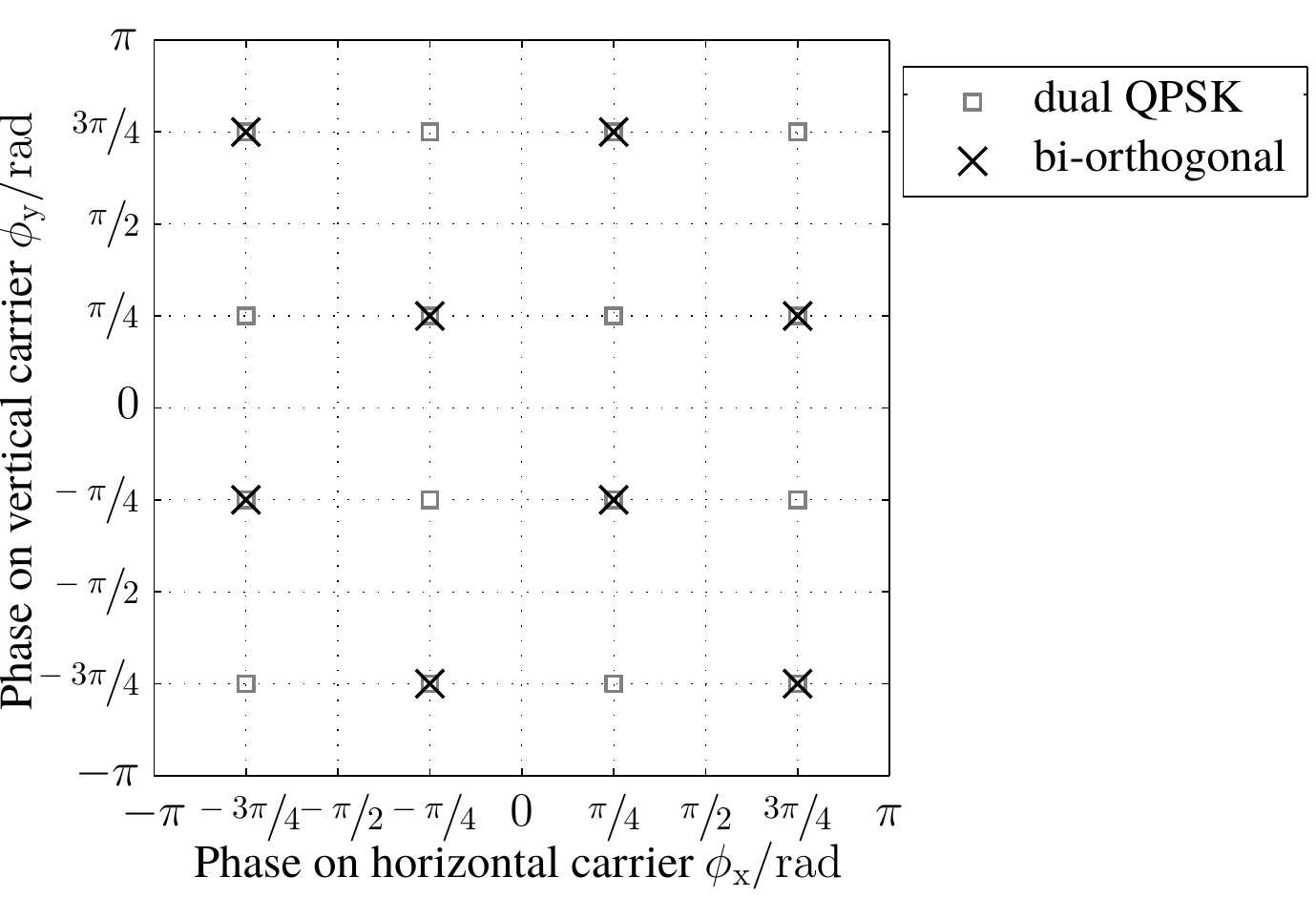}}
\caption{Phase diagram comparing dual QPSK and bi-orthogonal constellations}
\label{plot_bophase}
\end{figure}

By rotating this constellation in 4-D space it can be rendered into a constant-amplitude scheme, like the cylindrical constellation presented above \cite{Karlsson.2012}. The constellation diagram then looks like QPSK and a representation in \emph{phase space} is possible. This is shown in Figure \ref{plot_bophase}, comparing the constellation structure of dual, perfectly synchronous QPSK with that of bi-orthogonal signalling. For the bi-orthogonal constellation, half of the symbols are removed compared to the QPSK constellation. In this way, \SI{1}{\bps} of spectral efficiency is traded in for an increase in mutual distance by $\sqrt{2}$.

%
Figure \ref{plot_biortho} shows the simulation results comparing the symbol error rate of dual QPSK with that of the bi-orthogonal signal. The latter offers a power advantage of \SI{1,6}{\deci\bel} at an error rate of \num{1e-4}. This often accentuated comparison is questionable however because the spectral efficiency of the bi-orthogonal signal is significantly lower than that of QPSK. For this reason, a second simulation was run with a 3-point, PSK-like constellation arranged on the vertices of an equilateral triangle; in this way the spectral efficiency of the bi-orthogonal signal (3 bit per symbol) is approached very closely (\num{3,17} bit per symbol), legitimating a direct comparison. In this case the advantage of the bi-orthogonal signal shrinks to \SI{0,9}{\deci\bel}.
\subsection{Experimental demonstration of bi-orthogonal modulation over satellite}\label{ss_hardwareexp}
The bi-orthogonal constellation, due to its relatively simple structure, offers itself as a good choice for a demonstrative implementation of a 4-D modem. A transmitter and a receiver were implemented based on a programmable logic platform with analogue-to-digital and digital-to-analogue converters. A random symbol sequence at a rate of \SI{1,024}{\mega\baud} is transmitted. The signal is shaped with a root-raised-cosine pulse with roll-off $\alpha=\num{0,20}$ and output at an intermediate frequency of \SI{70}{\mega\hertz}. The receiver uses simple squaring for symbol recovery \cite{Franks.1974}. Carrier recovery is achieved by the QPSK carrier recovery method described by Sari in \cite{Sari.1988}.

\begin{figure}
\centering
\resizebox{\linewidth}{!}{\includegraphics{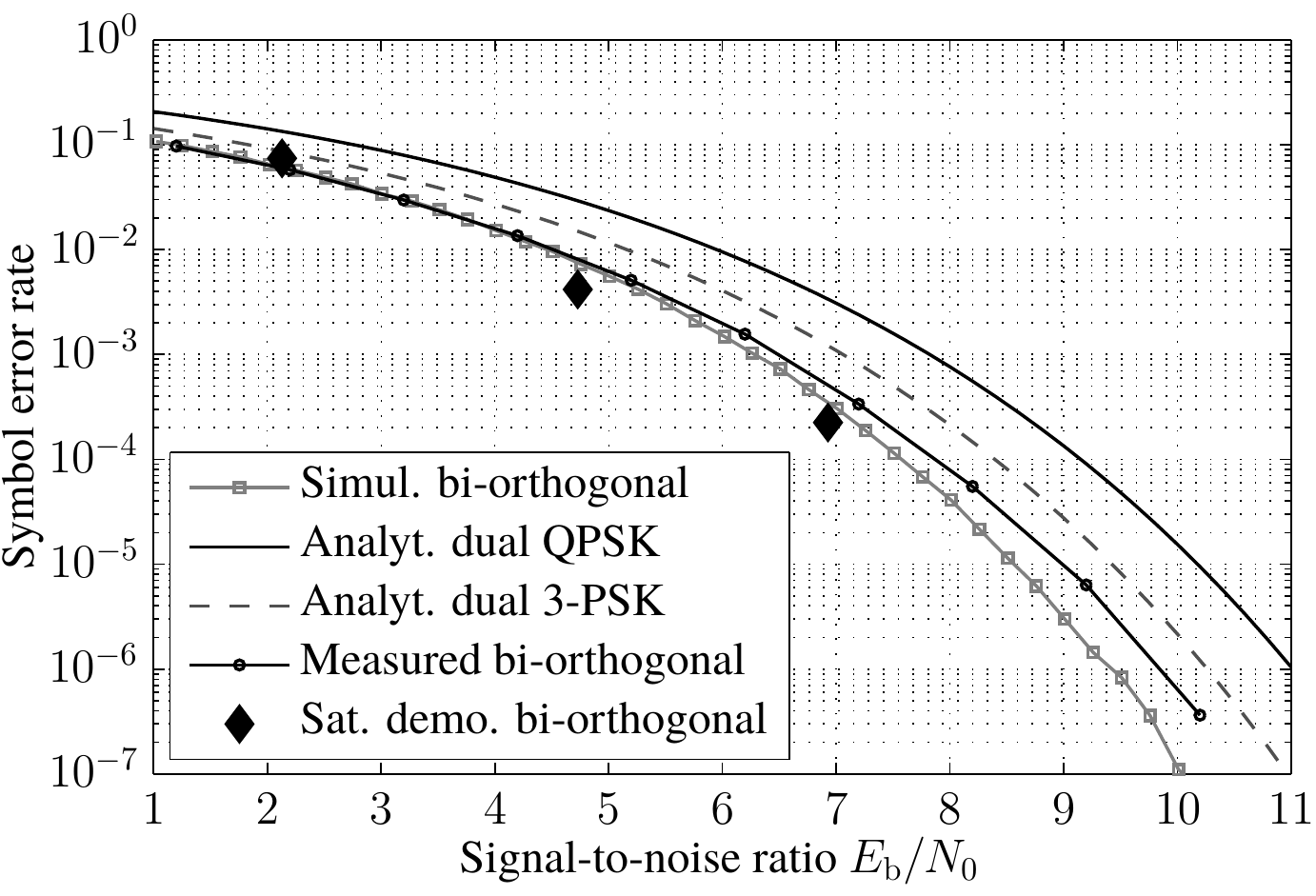}}
\caption{SER for bi-orthogonal modulation compared with QPSK.}
\label{plot_biortho}
\end{figure}

The implemented system is used to verify the claim of increased power efficiency in a practical setting. Different power levels of thermal noise are coupled into the signal path to synthesize a range of SNR steps. The symbol error rate is measured and the results are plotted in Figure \ref{plot_biortho}, alongside with the predictions. The experiment with the prototype verifies the predictions, even though a slight implementation loss becomes visible at higher SNR.

The same equipment was used in a satellite experiment, demonstrating the system's end-to-end functionality in its target setting. The purpose was to confirm the readiness of currently orbiting satellite transponders for 4-D signalling and to rule out any unforeseen impairments. The uplink was transmitted from an \SI{11}{\metre} ground station antenna at the SES site in Betzdorf, Luxembourg to the Astra 3B satellite. After frequency conversion, the signal was retransmitted and received by the same ground station. By varying the uplink power, it was possible to produce different SNR levels and measure the error rate as well.

The results from these measurements are displayed in Figure \ref{plot_biortho}, represented as diamonds. They confirm the previous on-ground measurements. As the noise power spectral density was measured at a different frequency, \SI{4}{\mega\hertz} higher than the actual carrier, a small bias manifests itself.
%
\subsection{Similarity to coding}
The methods used in the previous sections for constellation design come from coding theory. Even at a more fundamental level, the close relation between 4-D modulation and coding is evident: As a matter of fact, nothing prevents the creation of additional dimensions by just considering consecutive symbols in time and merging them into a higher-dimensional symbol. This is the essence of block-based error encoding. Also different frequency bands might be used to transmit more dimensions.

In this perspective, four-dimensional signalling can also be achieved by merging blocks of two consecutive symbols in a signal and soft-decoding them in the receiver. From a mathematically abstract viewpoint, there is no difference between these two approaches, when looking at spectral efficiency, bandwidth or power usage. However, different physical effects, advantages or constraints might play in favour or not of a polarization-, time- or frequency-based multidimensionality. An important physical advantage is an amplifier that can amplify both polarizations; this is illustrated in the next section.
\subsection{Peak-to-average power ratio}
Figure \ref{plot_256const} shows how constellation shaping increases the frequency of low-energy symbol-elements. Also the choice of the lattice structure, by expanding the constellation, increases the per-polarization peak-to-average power ratio (PAPR) of the carriers. 

Especially when each transponder is loaded with a single carrier, the amplifiers operate in or close to saturation for maximum output power. This distorts signals with high envelope variations; distortion which can be reduced by moving the amplifiers' operating point away from saturation, i.e.\ backing off. The price is lower output power, which in turn reduces the actual SNR. The SER to SNR curves shown in the previous section are oblivious of this, as $\sci{E}{b}$ designates only the \emph{average} power. The issue is less critical when multiple carriers are transmitted over the same transponder. Back-off is then mandatory by default and modulation types with higher PAPR may be used. 

A good heuristic to determine the amount of back-off required to maintain signal integrity is to look at the PAPR of the baseband signal. The PAPR of all proposed constellations was estimated from the waveforms generated during the simulations; they are shown in Table \ref{tab_papr}. The table is arranged in four columns: the two leftmost show the PAPR values computed from the symbols in the constellation. The two rightmost columns show the PAPR for pulse-shaped signals with roll-off $\alpha=\num{0,20}$, as yielded by simulations. Each of those two groups has a column for the \emph{combined} i.e.\ dual-polarization and the \emph{single}, per-carrier PAPR. The lines are ordered with decreasing spectral efficiency and also grouped accordingly.

Focussing on the pulse-shaped single-polarization column, the table allows the expected amount of required back-off to be compared. It shows that the PAPR of the large 4-D constellations (88-$\sci{D}{4}$-LAM, 256-$\sci{D}{4}$-LAM and 64-4D-PSK) is larger than that of the traditional constellations with which they are compared. By taking the PAPR \emph{ratio}, the required additional back-off can be roughly estimated. As an example, using 256-LAM, with an output back-off increased by $\num{6,6}/\num{5,6}=\SI{0,7}{\deci\bel}$ with respect to 16-QAM, results in a reduction of the power gain from \SI{1,3}{\deci\bel} down to \SI{0,6}{\deci\bel}. The same comparison with 16-APSK annihilates the gain expectation; this is true as well for the comparison of the 88-point LAM constellation with the hexagonal QAM. For the 64-4D-PSK constellation a net gain of \SI{0.3}{\deci\bel} remains.

Constant-amplitude constellations like QPSK or 8-PSK usually require no or only very little back-off in practical operation, despite the PAPR of around \num{3,5}. The same should be expected for the bi-orthogonal and the cylindrical 64-PSK constellation. Note that the PAPR of the 3-PSK is especially low, probably because symbol transitions do not pass through the origin.

The bi-orthogonal constellation was created by dividing the QPSK constellation into two complementary sets, of which one is removed (cf.~Figure \ref{plot_bophase}). In a variant of the bi-orthogonal signalling technique this set is flipped after every symbol. A first symbol would be mapped according to the crosses in Figure \ref{plot_bophase}, the following by the rectangles, the third by crosses again and so forth. This variant was simulated and is tabulated under \emph{bi-orthogonal alt.}. The combined PAPR of this waveform is the lowest of all simulated constellations. It seems that the modification reduces the number of symbol transitions through the origin in 4-D space.

The observation of low combined PAPR is valid for all of the proposed 4-D constellations: it is in all cases significantly lower than for the individual carriers. When looking at the pulse-shaped signals on the right side, this is even true for some of the classical constellations. If there was an amplifier able to operate on both polarizations, at the same carrier frequency simultaneously, the power constraint could be alleviated from
\begin{gather}
\sci{E}{s,x} < \sci{E}{max} \\
\sci{E}{s,y} < \sci{E}{max}
\label{eq_dualconstraints}
\end{gather}
to
\begin{gather}
\sci{E}{s,x}+\sci{E}{s,y} < 2\sci{E}{max}
\end{gather}
This would allow the use of 4-D constellations without the strong back-off penalty. The 64-4D-PSK constellation was designed with such a constraint in the design prescriptions and has a combined symbol PAPR of one. If such an amplifier could be manufactured and mounted on satellites, it would allow more efficient signalling in microwave broadband and satellite communications.
%
\begin{table}
\centering
\caption{Table of peak-to-average power ratios for proposed modulation techniques}
\label{tab_papr}
\begin{tabular}{ll r r r r}
\multicolumn{2}{l}{\textbf{Modulation}} & \multicolumn{2}{l}{\textbf{Symbols}} & \multicolumn{2}{l}{\textbf{RRCS }$\alpha=\num{0,20}$} \\ \hline
\multicolumn{2}{l}{ } & comb. & single & comb. & single \\  \hline
{} & 256-$\sci{D}{4}$-LAM & \num{1,3} & \num{2,5} & \num{3,6} & \num{6,6} \\
dual & 16-QAM & \num{1,8} & \num{1,8} & \num{4,3} & \num{5,6}\\
dual & 16-APSK & \num{1,3} & \num{1,3} & \num{3,3} & \num{4,0}\smallskip\\

{} & 88-$\sci{D}{4}$-LAM & \num{1,3} & \num{2,3} & \num{3,7} & \num{6,4} \\
{} & 64-4D-PSK & \num{1} & \num{1,9} & \num{3,0} & \num{5,5} \\
dual & 8-hex QAM & \num{1,6} & \num{1,6} & \num{4,2} & \num{5,3} \\

{} & hex.~cyl.~64-PSK & \num{1} & \num{1} & \num{3,0} & \num{3,4} \\
dual & 8-PSK & \num{1} & \num{1} & \num{3,0} & \num{3,5}\smallskip\\

{} & bi-orthogonal & \num{1} & \num{1} & \num{3,2} & \num{3,5} \\
{} & bi-orthogonal alt. & \num{1} & \num{1} & \num{2,5} & \num{3,5} \\
dual & QPSK & \num{1} & \num{1} & \num{3,0} & \num{3,5} \\
dual & 3-PSK & \num{1} & \num{1} & \num{2,8} & \num{2,9} 
\end{tabular}
\end{table}
\section{Dual-polarization synchronization}\label{s_synchro}
A satellite receiver is required to estimate the symbol clock and carrier phases and frequencies of an incoming signal. This is to adjust internal timing references and to synchronize itself to the signal using these estimated values, which is a prerequisite for successful reception. This is usually accomplished using feedback control systems called phase-locked loops (PLLs) \cite{Gardner.2005}.

Thermal noise in the signal will lead to fluctuating errors in these synchronization estimates, commonly called jitter. Jitter may be reduced by decreasing the loop bandwidth, i.e.\ performing the estimate over a longer measurement period. This however in turn reduces the synchronizer's capability to follow quick variations in the quantities to be estimated, like timing variations and phase noise. Therefore, the designer of a synchronization system needs to adjust the loop bandwidth to find a good compromise between the robustness against thermal noise and the tracking speed.

As already mentioned in Section \ref{s_chanmod}, equal symbol clocks on both polarizations are mandatory in four-dimensional signalling, such that both symbol-elements can be merged into the combined symbol for detection. Also, as carrier frequencies are equal, only one local oscillator may be used for frequency conversion, such that offsets and phase noise are equal on both polarizations. 

By exploiting these symmetric carrier properties and estimating over both polarizations, it is possible to reduce the thermal noise induced jitter without changing the measurement period. An improvement of synchronization fidelity should be expected in this way. To show this in theory, the signal model is modified to have equal carrier frequency offsets:
\begin{equation}
\Delta\sci{\omega}{x\vphantom{y}} = \Delta\sci{\omega}{y} = \Delta\omega
\end{equation}
It is realistic to assume that the time delay on both polarizations is not exactly the same because equal path-length is expensive to set up for a satellite channel. In that case, $\sci{\phi}{0}$ and $\tau$ would be offset by a fixed amount between the polarizations. Such a fixed offset can be estimated by a long-term average and remains without impact for the synchronizer. It will therefore be neglected in the following developments and the related time-dependent signal parameters can be assumed equal here as well:
\begin{gather}
\sci{\phi}{x,0} = \sci{\phi}{y,0} = \sci{\phi}{0}\\
\sci{\tau}{x} = \sci{\tau}{y} = \tau
\end{gather}
The signal definition from \eqref{eq_sxsy_full} can now be simplified:
\begin{equation}
\vc{s} = \shortexp{\j\left[\left(\omega+\Delta\omega\right) t + \sci{\phi}{0}\right]} \sum_{i} \begin{pmatrix} \sci{c}{x,i} \\ \sci{c}{y,i} \end{pmatrix} \cdot g\left(t-iT-\tau\right) \label{eq_sxy_sync}
\end{equation}
%
To assess the potential gain from using dual-polarization synchronization, the modified Cramer-Rao bound (MCRB) is used. The MCRB gives a lower bound for the achievable variance that can be expected from an unbiased estimate in a Gaussian noise process \cite{DAndrea.1994,Mengali.1997}. It is derived from the Cramer-Rao bound found in estimation theory, modified to ease the calculation for the present problem.

Starting from the log-likelihood function $\ln\Lambda\left(\lambda,\vc{u}\right)$ for the estimation of a constant parameter $\lambda$ in white Gaussian noise, extended to dual-polarization modulation, one can derive the expression for the MCRB by following along the lines of \cite{Mengali.1997}:
\begin{multline}
\ln\Lambda\left(\lambda,\vc{u}\right) = \\ \frac{-1}{2\sci{N}{0}}\int\limits_{\sci{T}{0}}\hm{\big( \vc{r}(t)-\vc{s}(t,\lambda,\vc{u})\big)}\big( \vc{r}(t)-\vc{s}(t,\lambda,\vc{u})\big)\,dt
\end{multline}
\begin{align}
\text{MCRB}(\lambda) = & \frac{1}{\Eptx{\vc{n},\vc{u}}{\left(\frac{\partial\,\ln\Lambda\left(\lambda,\vc{u}\right) }{\partial\lambda}\right)^2}} \label{eq_dualmcrb1}\\
= & \frac{-1}{\Eptx{\vc{n},\vc{u}}{\frac{\partial^2\,\ln\Lambda\left(\lambda,\vc{u}\right)}{\partial\lambda^2}}}\\
= & \frac{\sci{N}{0}}{\Eptx{\vc{u}}{\int\limits_{\sci{T}{0}}\hm{\left(\frac{\partial\,\vc{s}(t,\lambda,\vc{u})}{\partial \lambda}\right)}\left(\frac{\partial\,\vc{s}(t,\lambda,\vc{u})}{\partial \lambda}\right)dt}} \label{eq_dualmcrb2}
\end{align}
The vector $\vc{u}$ contains the parameters that are not to be estimated and are thus modelled as random, as well as the data symbols $\vci{c}{i}$. In \eqref{eq_dualmcrb1} the expected value is over the noise $\vc{n}$ and the vector $\vc{u}$, while in \eqref{eq_dualmcrb2} only $\vc{u}$ remains. The first equality is a known transformation for expected values exposed in \cite{vanTrees.2001}. Equation \eqref{eq_dualmcrb2} is obtained by performing the derivations and taking the partial expected value over the noise process $\vc{n}$.

For every of the three parameters $\Delta\omega$, $\sci{\phi}{0}$ and $\tau$, the MCRB resolves to the same expressions as given in \cite{Mengali.1997}. Indeed the MCRBs are the same as for the classic single-polarization case, except with the signal power doubled. In retrospective such a result might seem obvious as twice the signal energy is available to estimate the same quantity as in the single-polarization case. As a consequence, synchronizing using a dual-polarization signal has the potential to reduce the jitter variance by a factor two.

A similar reflection can also be made for a pilot-aided tracking system: If the quantity to be estimated is equal on both polarizations, twice the number of pilots are available to estimate the same quantity. Alternatively, one could just transmit pilots on one carrier, leaving the second exclusively for payload-data. In this case, the jitter variance would remain the same, but the efficiency of carrier usage would increase.

The MCRB only gives an indication on how well a best possible estimator would perform. To corroborate these analytical findings using a specific example, the variance of jitter in a timing recovery loop with a Gardner Error Detector (GED) \cite{Gardner.1986} is simulated. 16-QAM signals are used in the simulation, both in single- and dual-polarization operation; in this way both situations can easily be compared.

\begin{figure}
\centering
\rotatebox{0}{\resizebox{\linewidth}{!}{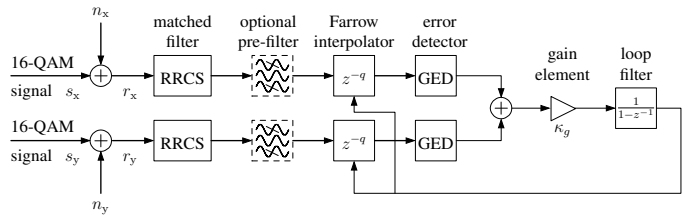}}
\caption{Block diagram for Gardner-Error-Detector simulations on a dual-polarization 16-QAM signal.}
\label{fig_GEDblock}
\end{figure}

A block diagram of the simulated system is shown in Figure \ref{fig_GEDblock}. Two noisy 16-QAM signals serve as inputs and are subsequently matched-filtered by a root-raised-cosine (RRCS). The timing recovery loop has two GEDs, one for each polarization. The error signals are added together and integrated to provide an input to the Farrow-interpolators \cite{Erup.1993}. An optional pre-filter can be used to reduce self-noise and thus allow better performance in high SNR \cite{DAndrea.1996}.

The performance of the simulated tracking system is compared with the normalized MCRB from \cite{Mengali.1997}:
\begin{equation}
\frac{1}{T^2}\times \text{MCRB}(\tau) = \frac{\sci{B}{N}T}{4\pi^2\zeta}\frac{\sci{N}{0}}{\sci{E}{s}}\label{eq_mcrbtau}
\end{equation}
$\sci{B}{N}$ is the loop noise bandwidth such that $1/2\sci{B}{N}T$ corresponds to the number of symbol periods measured \cite{Mengali.1997}. $\xi$ is a dimensionless parameter depending on the pulse-shape. It plays an important role as timing is easier to estimate with wideband pulses. The simulation used a roll-off value of $\alpha=\num{0,20}$, leading to $\xi = \num{0,852}$. As classic 16-QAM is used, the signal-to-noise ratio is expressed with $\sci{E}{s}/\sci{N}{0}$. To compute the MCRB for a dual-polarization signal, the symbol energy $\sci{E}{s}$ is simply doubled. 

\begin{figure}
\centering
\resizebox{\linewidth}{!}{\includegraphics{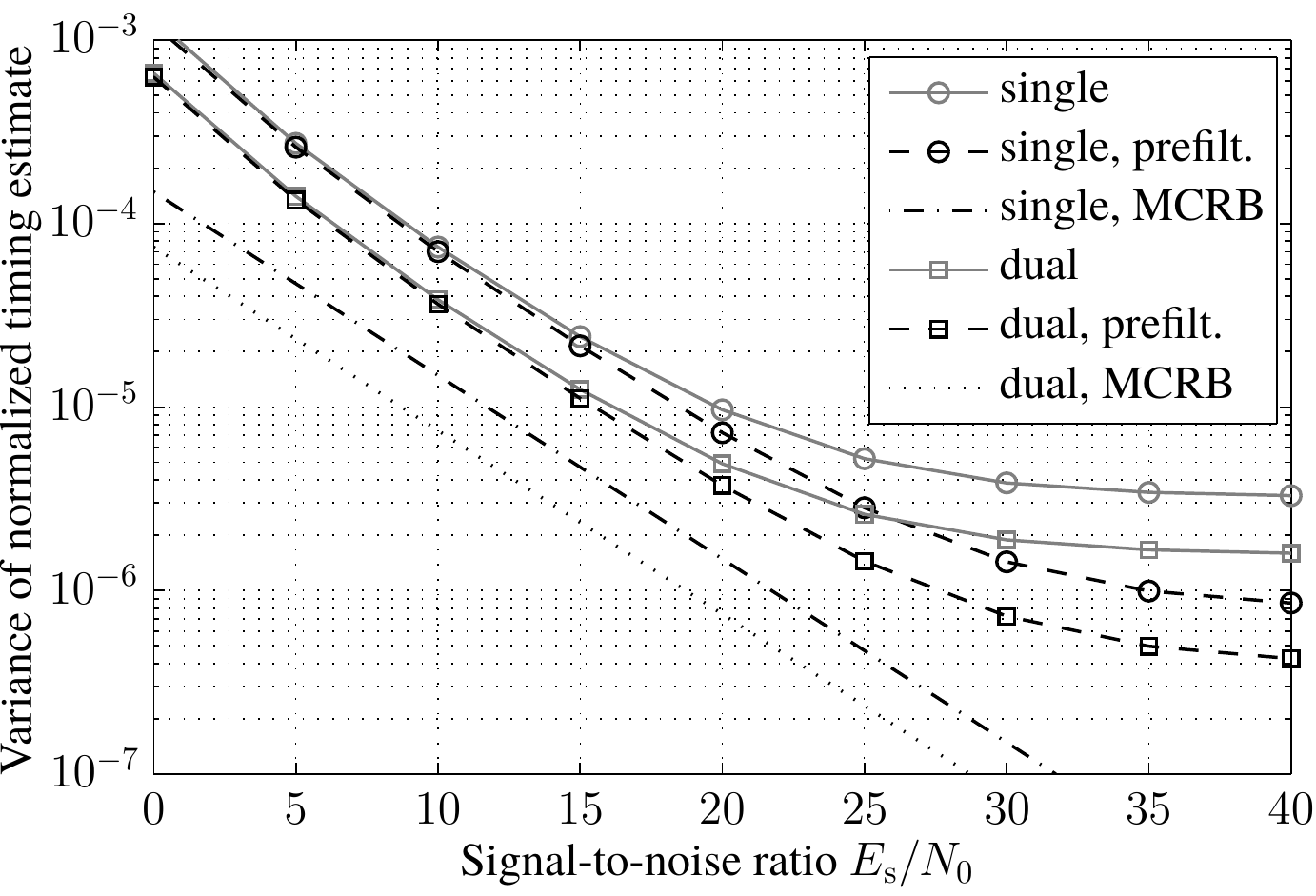}}
\caption{Variance of normalized GED timing estimate on 16-QAM signals with a roll-off $\alpha = \num{0,20}$ and a loop bandwidth $\sci{B}{N}T=\num{5e-4}$.}
\label{plot_ged020}
\end{figure}

Figure \ref{plot_ged020} shows the normalized timing variance as a function of $\sci{E}{s}/\sci{N}{0}$ for the simulated single- and dual-polarization signals with and without pre-filtering as well as the corresponding MCRB.

In the low-SNR region, the GED tracking loop jitter variance is higher than the MCRB by a factor of \num{5,6} for both cases. Self-noise limits the performance in high SNR, a phenomenon that the pre-filters are able to mitigate somewhat. According to \cite{Mengali.1997}, the MCRB implicitly assumes known data symbols and is usually closely approached by decision- and data-directed synchronization methods. The simulations confirm that it is too optimistic for this non-data-aided method. More importantly, in both cases with and without pre-filter, the jitter is reduced by half for all noise levels in the dual-polarization setup. Even the asymptotic limits caused by self-noise are pushed down by $\sfrac{1}{2}$, such that dual-polarization tracking outperforms pre-filtered, single-polarization tracking until up to a SNR of about \SI{25}{\deci\bel}, which includes most practical cases.

It can be concluded that a 4-D receiver can increase its tracking precision by estimating the symbol clock over both polarizations. Judging from the respective MCRBs, the same should also be true for carrier and phase recovery. This is especially useful in satellite communications, where the transponder transmit power is usually limited by physical or regulatory constraints. For this reason the loop bandwidth is often the only tunable parameter to optimize synchronizers. If conflicting requirements do not allow a satisfactory solution, dual-polarization synchronization can be a solution.
%
%
%
%
\section{Conclusions}\label{s_concl}
The signal model for four-dimensional modulation over satellite was described. A proven tool for constellation design was reviewed and used to generate different four-dimensional constellations for the satellite link.

Numerical computations and simulations showed these constellations to have a better power efficiency than comparable single-polarization techniques. The increased envelope variations resulting from these techniques reduce this advantage when transponders are operated close to saturation, like in a single-carrier per transponder setup. Except from the presented bi-orthogonal signalling, they are preferably used in multiple-carrier applications, where amplifier back-off is mandatory and no additional losses are expected. For the bi-orthogonal signal a comparison with a constellation on an equilateral triangle was proposed. The simulation results are backed by an experimental verification using an implemented demonstration modem with the bi-orthogonal signal. The experiments were concluded by demonstrating a successful transmission over satellite.

An important aspect is the relationship between coding and four-dimensional modulation. The conclusion is that a real physical advantage must be present to exploit the advantages of a dual-polarization signal; otherwise the exercise is of purely mathematical nature.

One such advantage is the improved tracking accuracy that can be obtained in the receivers by adding the error signals from two error detectors, one on each polarization. Simulations on a timing-loop using a Gardner timing-error detector confirmed the respective predictions.

Four-dimensional signalling over satellite can be regarded as a real alternative to the previously used single-polarization techniques. It offers improved tracking, more flexibility in the bandwidth usage, electronic cross-polarization compensation and the signalling gain from improved constellations. On one hand, each of these advantages by itself might not be decisive. On the other hand, implementing the capability for 4-D modulation comes at a very low additional cost; subsequently, all the advantages are available together. In addition, a system capable of four-dimensional signalling can always be operated in a backwards-compatible way with classical 2-D symbol constellations.

From a practical engineering point of view, only little changes need to be made to existing systems in order to profit from the advantages of four-dimensional modulation. Ground stations and satellite transponders are usually perfectly capable of transmitting on two polarizations. On transponders from the digital era channels are already aligned between polarizations. In order to also profit from the tracking benefit, system design needs to ensure that both polarizations use the same local oscillator for frequency conversion. The readiness of the space-segment was demonstrated by the live satellite demonstration; what remains is the reception equipment on the consumer side. Consumer-grade low-noise block converters (LNBs) are already capable of downconverting both polarizations simultaneously and coherently (cf.\ Twin LNB \cite{SESS.A..2007TWIN}). Only the tuners in the reception equipment need to be adapted to provide for two inputs.

Some effort should be put into the search for densely-packed four-dimensional constellations with APSK-like constellation diagrams. An immense advantage could also be provided by an amplifier able to amplify both polarizations together. It may be possible to develop such a capability from gyrotrons or multi-port amplifiers.
\appendix
\section{The 8-point hexagonal constellation}\label{apx_hexqam}
The coordinates of the points in the hexagonal constellation from Section \ref{ss_lam} are stated hereunder:
\begin{multline}
\frac{d}{2}\cdot\left\{\begin{pmatrix} -1 \end{pmatrix},\,
\begin{pmatrix} 1 \end{pmatrix},\,
\begin{pmatrix} -\j\sqrt{3} \end{pmatrix},\,
\begin{pmatrix} \j\sqrt{3} \end{pmatrix},\,
\begin{pmatrix} -2-\j\sqrt{3} \end{pmatrix},\right.\\
\left.\begin{pmatrix} -2+\j\sqrt{3} \end{pmatrix},\,
\begin{pmatrix} 2-\j\sqrt{3} \end{pmatrix},\,
\begin{pmatrix} 2+\j\sqrt{3} \end{pmatrix}\right\}
\end{multline}
\section*{Acknowledgement}
The present project is supported by the National Research Fund, Luxembourg. The work is conducted in a partnership between the University of Luxembourg and SES S.A.\ .
\renewcommand*{\UrlFont}{\rmfamily}
\printbibliography
\begin{IEEEbiography}[{\includegraphics[width=1in,height=1.25in,clip,keepaspectratio]{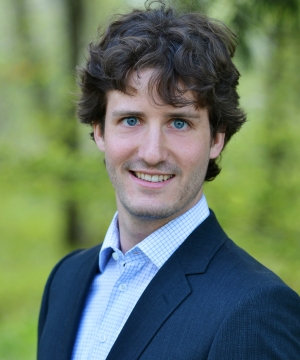}}]{Lionel Arend,}
born 1986 in Luxembourg, received his Electrical Engineering degree (Dipl.-Ing.) from the Technical University in Munich in 2011. In 2015, he was awarded the Ph.D.~degree by the University of Luxembourg for his thesis on combined-polarization modulation, in a partnership project together with the satellite operator SES. Lionel's scientific interests include modulation methods and digital signal processing in satellite communications as well as phase-locked loops and recovery techniques for receivers.
\end{IEEEbiography}
\begin{IEEEbiography}[{\includegraphics[width=1in,height=1.25in,clip,keepaspectratio]{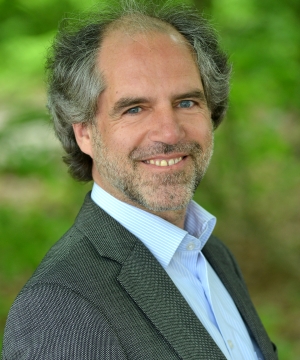}}]{Jens Krause}
was born in Werdohl, Germany, in 1963. He received the Dipl.-Ing. degree in 1987 and the Ph.D.~degree in 1993, both in electrical engineering from Universität Karlsruhe, Germany. He has held a scientific employee position at the Institute for Communications Technology, University of Karlsruhe, from 1988 to 1993. From 1994 to 1996 he has been a R\&D engineer in the CATV department of Richard Hirschmann GmbH in Germany. Since 1996 he works at the satellite operator SES S.A. in Luxembourg, with the current position in systems engineering. He represents SES in standardization organisations including ETSI and DVB. His technical interests include satellite communications in general, modulation and coding, satellite antennas, technical standardization.
\end{IEEEbiography}
\begin{IEEEbiography}[{\includegraphics[width=1in,height=1.25in,clip,keepaspectratio]{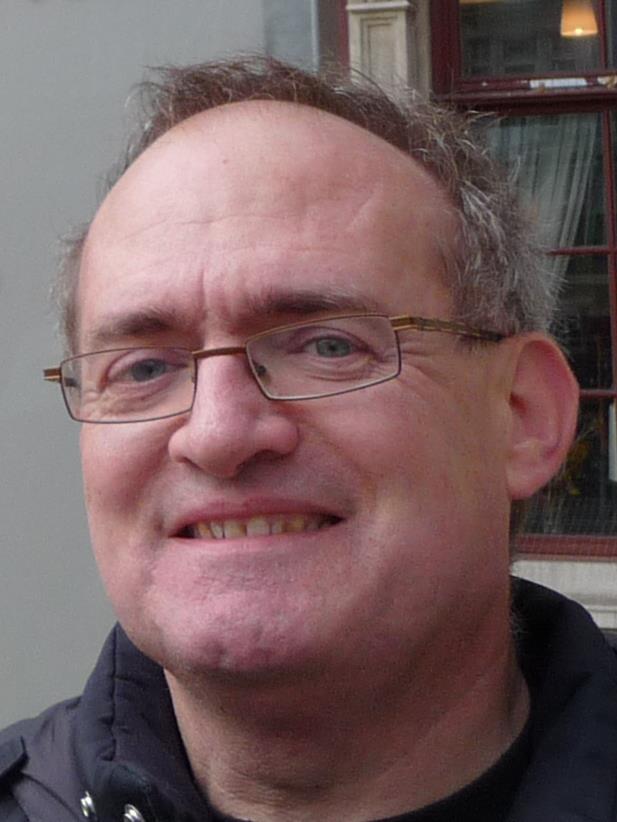}}]{Michel Marso}
was born in Luxembourg, in 1960. He received the Dipl.-Ing. and Ph.D.~degrees from the Technical University (RWTH), Aachen, Germany in 1987 and 1991, respectively. Between 1991 and 2008 he was with the Institute of Thin Films and Interfaces at the Research Center in Jülich, Germany, where he worked on GaN based transistors and nanodevices and on LT-GaAs based photodetectors for THz-generation. Since 2008, he is professor at the University of Luxembourg. His research interests center on semiconductor devices, generation and application of submillimeter radiation and on novel antenna concepts.
\end{IEEEbiography}
\begin{IEEEbiography}[{\includegraphics[width=1in,height=1.25in,clip,keepaspectratio]{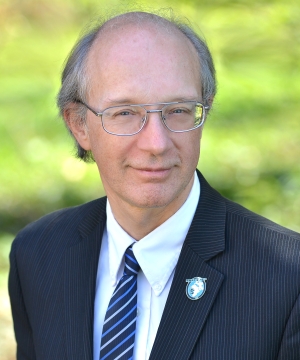}}]{Ray Sperber}
is in the SES Engineering Planning Department, and defined the initial "Polarization Signalling" project. He has worked in the space industry since 1978, mostly in various domains of satellite communications. He obtained a BS in Physics in 1976, a MS in Aeronautics \& Astronautics in 1978, both from MIT, and an MBA from Sacred Heart Luxembourg in 1997. His research interests are optics, communications and value of information.
\end{IEEEbiography}
\end{document}

%% file: f/GED_blockdiagram_LF.pdf_t
\begin{picture}(0,0)%
\includegraphics{GED_blockdiagram_LF.pdf}%
\end{picture}%
\setlength{\unitlength}{4144sp}%
\begingroup\makeatletter\ifx\SetFigFont\undefined%
\gdef\SetFigFont#1#2#3#4#5{%
  \reset@font\fontsize{#1}{#2pt}%
  \fontfamily{#3}\fontseries{#4}\fontshape{#5}%
  \selectfont}%
\fi\endgroup%
\begin{picture}(7414,2421)(2679,-1219)
\put(9586,299){\makebox(0,0)[b]{\smash{{\SetFigFont{12}{14.4}{\familydefault}{\mddefault}{\updefault}{\color[rgb]{0,0,0}filter}%
}}}}
\put(8776,299){\makebox(0,0)[b]{\smash{{\SetFigFont{12}{14.4}{\familydefault}{\mddefault}{\updefault}{\color[rgb]{0,0,0}element}%
}}}}
\put(9586,-61){\makebox(0,0)[b]{\smash{{\SetFigFont{12}{14.4}{\familydefault}{\mddefault}{\updefault}{\color[rgb]{0,0,0}$\frac{1}{1-z^{-1}}$}%
}}}}
\put(8776,-286){\makebox(0,0)[b]{\smash{{\SetFigFont{12}{14.4}{\familydefault}{\mddefault}{\updefault}{\color[rgb]{0,0,0}$\kappa_g$}%
}}}}
\put(7381,-466){\makebox(0,0)[b]{\smash{{\SetFigFont{12}{14.4}{\familydefault}{\mddefault}{\updefault}{\color[rgb]{0,0,0}GED}%
}}}}
\put(7381,254){\makebox(0,0)[b]{\smash{{\SetFigFont{12}{14.4}{\familydefault}{\mddefault}{\updefault}{\color[rgb]{0,0,0}GED}%
}}}}
\put(6481,254){\makebox(0,0)[b]{\smash{{\SetFigFont{12}{14.4}{\familydefault}{\mddefault}{\updefault}{\color[rgb]{0,0,0}$z^{-q}$}%
}}}}
\put(6481,-466){\makebox(0,0)[b]{\smash{{\SetFigFont{12}{14.4}{\familydefault}{\mddefault}{\updefault}{\color[rgb]{0,0,0}$z^{-q}$}%
}}}}
\put(6481,659){\makebox(0,0)[b]{\smash{{\SetFigFont{12}{14.4}{\familydefault}{\mddefault}{\updefault}{\color[rgb]{0,0,0}interpolator}%
}}}}
\put(5536,659){\makebox(0,0)[b]{\smash{{\SetFigFont{12}{14.4}{\familydefault}{\mddefault}{\updefault}{\color[rgb]{0,0,0}pre-filter}%
}}}}
\put(4591,254){\makebox(0,0)[b]{\smash{{\SetFigFont{12}{14.4}{\familydefault}{\mddefault}{\updefault}{\color[rgb]{0,0,0}RRCS}%
}}}}
\put(4591,-466){\makebox(0,0)[b]{\smash{{\SetFigFont{12}{14.4}{\familydefault}{\mddefault}{\updefault}{\color[rgb]{0,0,0}RRCS}%
}}}}
\put(4591,659){\makebox(0,0)[b]{\smash{{\SetFigFont{12}{14.4}{\familydefault}{\mddefault}{\updefault}{\color[rgb]{0,0,0}filter}%
}}}}
\put(3691,1019){\makebox(0,0)[b]{\smash{{\SetFigFont{12}{14.4}{\familydefault}{\mddefault}{\updefault}{\color[rgb]{0,0,0}$\sci{n}{x}$}%
}}}}
\put(3691,-1141){\makebox(0,0)[b]{\smash{{\SetFigFont{12}{14.4}{\familydefault}{\mddefault}{\updefault}{\color[rgb]{0,0,0}$\sci{n}{y}$}%
}}}}
\put(7426,659){\makebox(0,0)[b]{\smash{{\SetFigFont{12}{14.4}{\familydefault}{\mddefault}{\updefault}{\color[rgb]{0,0,0}detector}%
}}}}
\put(4096,119){\makebox(0,0)[rb]{\smash{{\SetFigFont{12}{14.4}{\familydefault}{\mddefault}{\updefault}{\color[rgb]{0,0,0}$\sci{r}{x}$}%
}}}}
\put(2701,119){\makebox(0,0)[lb]{\smash{{\SetFigFont{12}{14.4}{\familydefault}{\mddefault}{\updefault}{\color[rgb]{0,0,0}signal}%
}}}}
\put(4096,-601){\makebox(0,0)[rb]{\smash{{\SetFigFont{12}{14.4}{\familydefault}{\mddefault}{\updefault}{\color[rgb]{0,0,0}$\sci{r}{y}$}%
}}}}
\put(2701,-601){\makebox(0,0)[lb]{\smash{{\SetFigFont{12}{14.4}{\familydefault}{\mddefault}{\updefault}{\color[rgb]{0,0,0}signal}%
}}}}
\put(3466,-601){\makebox(0,0)[rb]{\smash{{\SetFigFont{12}{14.4}{\familydefault}{\mddefault}{\updefault}{\color[rgb]{0,0,0}$\sci{s}{y}$}%
}}}}
\put(3466,119){\makebox(0,0)[rb]{\smash{{\SetFigFont{12}{14.4}{\familydefault}{\mddefault}{\updefault}{\color[rgb]{0,0,0}$\sci{s}{x}$}%
}}}}
\put(2701,-286){\makebox(0,0)[lb]{\smash{{\SetFigFont{12}{14.4}{\familydefault}{\mddefault}{\updefault}{\color[rgb]{0,0,0}16-QAM}%
}}}}
\put(2701,434){\makebox(0,0)[lb]{\smash{{\SetFigFont{12}{14.4}{\familydefault}{\mddefault}{\updefault}{\color[rgb]{0,0,0}16-QAM}%
}}}}
\put(4591,884){\makebox(0,0)[b]{\smash{{\SetFigFont{12}{14.4}{\familydefault}{\mddefault}{\updefault}{\color[rgb]{0,0,0}matched}%
}}}}
\put(5536,884){\makebox(0,0)[b]{\smash{{\SetFigFont{12}{14.4}{\familydefault}{\mddefault}{\updefault}{\color[rgb]{0,0,0}optional}%
}}}}
\put(6481,884){\makebox(0,0)[b]{\smash{{\SetFigFont{12}{14.4}{\familydefault}{\mddefault}{\updefault}{\color[rgb]{0,0,0}Farrow}%
}}}}
\put(7426,884){\makebox(0,0)[b]{\smash{{\SetFigFont{12}{14.4}{\familydefault}{\mddefault}{\updefault}{\color[rgb]{0,0,0}error}%
}}}}
\put(8776,524){\makebox(0,0)[b]{\smash{{\SetFigFont{12}{14.4}{\familydefault}{\mddefault}{\updefault}{\color[rgb]{0,0,0}gain}%
}}}}
\put(9586,524){\makebox(0,0)[b]{\smash{{\SetFigFont{12}{14.4}{\familydefault}{\mddefault}{\updefault}{\color[rgb]{0,0,0}loop}%
}}}}
\end{picture}%